\documentclass[
preprint,
amsmath,amssymb,
aps,
]{revtex4-2}

\usepackage{float}
\usepackage{graphicx}
\usepackage{multirow}
\usepackage{tabularx}

\usepackage{dcolumn}
\usepackage{bm}

\RequirePackage[colorlinks=true
  ,urlcolor=blue
  ,anchorcolor=blue
  ,citecolor=blue
  ,filecolor=blue
  ,linkcolor=blue
  ,menucolor=blue
  ,linktocpage=true
  ,pdfproducer=medialab
  ,pdfa=true
]{hyperref}

\usepackage[dvipsnames]{xcolor}
\usepackage{comment}
\usepackage{mathtools}
\usepackage{xcolor}
\usepackage{bigstrut}
\usepackage{booktabs}
\usepackage{array}

\usepackage[compat=1.0.0]{tikz-feynman}

\newcommand{\be}{\begin{equation}}
\newcommand{\ee}{\end{equation}}
\newcommand{\beq}{\begin{equation}}
\newcommand{\eeq}{\end{equation}}
\def\bsp#1\esp{\begin{split}#1\end{split}}
\def\bal#1\eal{\begin{align}#1\end{align}}
\newcommand{\bea}{\begin{eqnarray}}
\newcommand{\eea}{\end{eqnarray}}

\newcommand{\ri}{\text{i}} 
\newcommand{\rc}{\text{c}}

\newcommand{\rL}{\text{L}} 

\newcommand{\tZ}{\theta_Z} 

\let\oldref\ref
\renewcommand{\ref}[1]{(\oldref{#1})} 
\numberwithin{equation}{section}

\newcommand\tS   {\theta_S}
\newcommand{\tG}{\theta_G}

\newcommand\MeV  {\ensuremath{\mathrm{MeV}}}
\newcommand\GeV  {\ensuremath{\mathrm{GeV}}}
\newcommand\TeV  {\ensuremath{\mathrm{TeV}}}

\newcommand\rs   {\ensuremath{\mathrm{s}}}

\newcommand\cL   {\ensuremath{\mathcal{L}}}
\newcommand\cz {\ensuremath{ \rc_Z}}
\newcommand\sz {\ensuremath{ \rs_Z}}

\newcommand\cs {\ensuremath{ \rc_S}}
\newcommand\sss {\ensuremath{ \rs_S}}
\newcommand\tanb {\ensuremath{ \tan\beta}}
\makeatletter
\newcommand{\undersim}[1]{\mathrel{\mathpalette\@undersim{#1}}}
\newcommand{\@undersim}[2]{%
  \vcenter{%
    \ialign{%
      ##\cr
      $\m@th#1#2$\cr
      \noalign{\nointerlineskip\kern.2ex}
      $\m@th#1\sim$\cr
      \noalign{\kern-.4ex}
    }%
  }%
}

\makeatother

\begin{document}
\title{Precision Higgs Constraints in U(1) Extensions of the Standard Model with a Light Z'-Boson}

\author{Zolt\'an P\'eli}
\email{zoltan.peli@ttk.elte.hu}
\affiliation{Institute for Theoretical Physics, ELTE E\"otv\"os Lor\'and University,
P\'azm\'any P\'eter s\'et\'any 1/A, 1117 Budapest, Hungary
}
\author{Zolt\'an Tr\'ocs\'anyi}
\email{zoltan.trocsanyi@cern.ch}
\affiliation{Institute for Theoretical Physics, ELTE E\"otv\"os Lor\'and University and
HUN-REN ELTE Theoretical Physics Research Group,
P\'azm\'any P\'eter s\'et\'any 1/A, 1117 Budapest, Hungary, also at\\
University of Debrecen, Bem t\'er 18/A, 4026 Debrecen, Hungary
}

\begin{abstract}
Anomaly free $U(1)$ extensions of the standard model (SM) predict a new 
neutral gauge boson $Z'$. When the $Z'$ obtains its mass from the 
spontaneous breaking of the new $U(1)$ symmetry by a new complex scalar field, 
the model also predicts a second real scalar $s$ and the searches for the new 
scalar and the new gauge boson become intertwined. We present the computation 
of production cross sections and decay widths of such a scalar $s$ in models 
with a light $Z'$ boson, when the decay $h\to Z' Z'$ may have a sizeable 
branching ratio. We show how Higgs signal strength measurement in this 
channel can provide stricter exclusion bounds on the parameters of the model
than those obtained from the total signal strength for Higgs boson production.
\end{abstract}

\maketitle

\section{Introduction}

While the discovery of the Higgs boson \cite{Aad:2012tfa,Chatrchyan:2012xdj}
has established the existence of a scalar elementary particle, the thorough 
understanding of the role of scalar fields in Nature remains elusive. So far 
all experimental results are in agreement with the structure of the scalar 
sector of the standard model (SM) \cite{ParticleDataGroup:2024cfk}, although the scalar 
potential has not yet been fully confirmed experimentally, which allows for 
the existence of an extended scalar sector. Indeed, there is a vigorous 
experimental search for new scalar particles at the LHC \cite{CMS:2024phk}. 
The more complex such an extended sector, the more new particles should exist 
and the more difficult the search strategies.

\begin{figure}[h!]
    \centering
    \begin{tikzpicture}
        \begin{feynman}
            \vertex (h0) at (-2, 0) {$h$}; 
            \vertex (h) at (0, 0); 
            \vertex (z1) at (2, 1) {$Z'$}; 
            \vertex (z2) at (2, -1) {$Z'$}; 

            \diagram* {
                (h0) -- [scalar] (h),
                (h) -- [photon] (z1), 
                (h) -- [photon] (z2)  
            };
        \end{feynman}
    \end{tikzpicture}
    \caption{Tree-level diagram of the Higgs boson decaying into two $Z'$ bosons.}
    \label{fig:higgs_to_zpzp}
\end{figure}
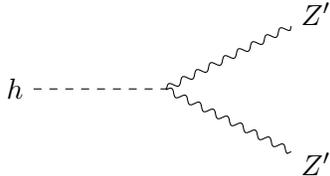

The $U(1)$ extensions of the SM have the potential to explain several
beyond the standard model (BSM) phenomena at the cost of predicting the existence of $Z'$, a new
neutral gauge boson. In the simplest scenario $Z'$ acquires its mass
from the spontaneous breaking of a new scalar field $\chi$ \cite{Appelquist:2002mw},  
hence, the model also predicts a new scalar particle $s$. 
$Z'$ bosons appear in a wide variety of models, for a comprehensive review on them see Ref.~
\cite{Langacker:2008yv}.
Experiments 
searched extensively for new scalar particles, 
as well as a new $Z'$ boson (for an incomplete list we refer to 
Refs.~\cite{Robens:2022cun,ATLAS:2019erb,CMS:2021ctt,NA64:2019imj}). In such $U(1)$ extensions the
existence of $s$ and $Z'$ are interconnected.  For instance, if $Z'$ is
sufficiently light, then the Higgs boson and the new scalar can decay
into a $Z'$ pair 
as shown on Figure.~\ref{fig:higgs_to_zpzp}.
The channel $h \to Z' Z'$ can significantly
alter the decay properties of the Higgs boson, which consequently can
be used to constrain the free parameters of the model.


An example of such a simple $U(1)$ extension of the standard model,
with the potential to explain all established experimental
observations that cannot be interpreted within the SM, is the
superweak extension of the standard model (SWSM) \cite{Trocsanyi:2018bkm}.
It is designed to explain (i) the non-vanishing masses of neutrinos, 
(ii) the nature of dark matter,
(iii) the origin of baryon asymmetry, and 
(iv) the role of scalar fields in the Universe.

The SWSM extends the gauge group of the SM
$G = SU(3)_\rc \otimes SU(2)_\rL \otimes U(1)_Y$, to $G \otimes U(1)_z$.
The corresponding neutral gauge field becomes massive by introducing a
complex scalar field $\chi$ with non-zero vacuum expectation value (VEV).
The gauge and gravity anomalies are required to cancel, which is achieved 
by introducing three generations of right-handed neutrinos -- dubbed 
heavy neutral lepton, (HNL) below -- with properly chosen $z$ charges $z_N$. 
If we also allow the presence of Dirac-type Yukawa terms for the HNLs, 
then two independent $z$ charges remain. We choose the $z$ charge $z_\phi$ 
of the Brout-Englert-Higgs (BEH) field and $z_N$ to be the free, model 
dependent parameters.
All new fields are neutral under the standard model gauge interactions. 
The SWSM fixes the $z$ charge of the BEH field to $z_\phi = 1$, 
and those of the HNLs to $z_N = \frac12$ by allowing also Majorana-type
Yukawa term in the Lagrangian for the HNLs and normalizing the
new gauge coupling $g_z$ such that the new scalar has $z$ charge
$z_\chi = -1$. 

In this work we compute benchmark points for production and decay of a new 
scalar $s$ for models with a light $Z'$ boson. If the $Z'$ is light 
the decay $h\to Z' Z'$ may have a sizeable branching ratio
with dominant contribution independent of $z_\phi$ and $z_N$.
We show that exclusion limits on new singlet scalar particles obtained from 
signal strength measurements can be improved by taking into account the 
process $h\to Z' Z'$. We provide a {\em Mathematica} notebook that can be 
used flexibly to compute benchmark points relevant for scalar searches 
at the LHC. While the specific model we have in mind is the SWSM,
yet our analysis is valid for an arbitrary assignment of $z$ charges.

Experiments often search for the so-called dark photon, which 
appears in the $U(1)$ extensions with $z_\phi = z_N = 0$. These results 
can be translated to $U(1)$ extensions with arbitrary $z$ charge assignment
\cite{Ilten:2018crw}. The prospects of discovering a dark photon were 
studied extensively \cite{Curtin:2014cca} when one also takes into account the
channels $h\to Z Z'$ and $h\to Z'Z'$. Our approach is different in the sense
that we investigate the prospects of discovering a new scalar $s$ when the 
model also predicts a $Z'$ boson with unconstrained $z_\phi$ and $z_N$ charges.

\section{Scalar sector of the SWSM}
\label{sect:u1_extension}

In this section we collect the details of the model only to the extent 
used in the present analyses.

In the scalar sector, in addition to the $SU(2)_\rL$-doublet 
Brout-Englert-Higgs field
\begin{equation}
\phi=\left(\!\!\begin{array}{c}
               \phi^{+} \\
               \phi^{0}
             \end{array}\!\!\right) = 
\frac{1}{\sqrt{2}}
\left(\!\!\begin{array}{c}
          \phi_{1}+\ri\phi_{2} \\
          \phi_{3}+\ri\phi_{4}
        \end{array}\!\!
\right)
\,,
\end{equation}
the model contains a complex scalar SM singlet $\chi$.
The Lagrangian of the scalar fields contains the potential energy
\beq
\bsp
V(\phi,\chi) &= - \mu_\phi^2 |\phi|^2 - \mu_\chi^2 |\chi|^2
+ \left(|\phi|^2, |\chi|^2\right)
\left(\!\!\begin{array}{cc}
 \lambda_\phi & \frac{\lambda}2 \\ \frac{\lambda}2 & \lambda_\chi
\end{array}\!\!\right)
\left(\!\!\begin{array}{c}
|\phi|^2 \\ |\chi|^2
\end{array}\!\!\right) 
\subset  -\cL
\label{eq:V}
\esp
\eeq
where $|\phi|^2 =|\phi^+|^2 + |\phi^0|^2$.
After spontaneous symmetry breaking, we parametrize the scalar fields as
\beq
\bsp
\phi =\frac{1}{\sqrt{2}}\binom{-\ri \sqrt{2}\sigma^+}{v+h'+ \ri\sigma_\phi}
\,,\quad
\chi = \frac{1}{\sqrt{2}}(w + s' + \ri\sigma_\chi)
\esp
\eeq
where $v$ and $w$ are the vacuum expectation values of $\phi$ and $\chi$.
The fields $h'$ and $s'$ are real scalars, $\sigma^+$ is a charged, while 
$\sigma_\phi$ and $\sigma_\chi$ are neutral Goldstone bosons that are gauge eigenstates. 

The gauge and mass eigenstates are related by the rotations
\beq
\binom{h}{s} = \textbf{Z}_S
\binom{h'}{s'}
\,,\quad
\binom{\sigma_Z}{\sigma_{Z'}} = \textbf{Z}_G
\binom{\sigma_\phi}{\sigma_\chi}
\,,
\eeq
with
\beq
\textbf{Z}_X=
\begin{pmatrix}
\cos\theta_X & -\sin\theta_X
\\
\sin\theta_X & ~~ \cos\theta_X
\end{pmatrix}
\eeq
where we denoted the mass eigenstates with $h$, $s$ and $\sigma_Z$,
$\sigma_{Z'}$. The angles $\tS$ and $\tG$ 
are the scalar and Goldstone mixing angles that can
be determined by the diagonalization of the mass matrix of the real scalars 
and that of the neutral Goldstone bosons. In the following, we are going to use 
the abbreviations ${\rm c}_X = \cos\theta_X$ and ${\rm s}_X = \sin\theta_X$ for mixing angles.

\subsection{Scalar couplings}

The vertices that involve the scalars are related to the corresponding vertices
in the SM by simple proportionality factors involving the scalar mixing angle
$\tS$. Hence, if possible, we present the Feynman rules expressed using the
corresponding SM rule.
\begin{itemize}
\item Scalar-fermion couplings:
\begin{equation}
\Gamma_{hf\bar{f}} = \cs \Gamma_{hf\bar{f}}^{\text{SM}}
\,,\quad
\Gamma_{sf\bar{f}} = \sss \Gamma_{hf\bar{f}}^{\text{SM}}.
\label{eq:VSff}
\end{equation}
\item Scalar-vector boson couplings: ${\rm i}\Gamma_{SVV} g^{\mu\nu}$ where 
\begin{equation}
\Gamma_{hWW} = \cs \Gamma_{hWW}^{\text{SM}}
\,,\quad
\Gamma_{sWW} = \sss \Gamma_{hWW}^{\text{SM}}
\end{equation}
and
\begin{equation}
\bsp
\Gamma_{hZZ} &= 2 M_Z^2 \biggl(\cs \cz^2 - \frac{\sss}{\tanb}\sz^2 \biggr)\,,
\quad
\Gamma_{sZZ} = 2 M_Z^2  \biggl(\sss \cz^2 + \frac{\cs}{\tanb} \sz^2\biggr)\,,
\\
\Gamma_{hZZ'} &= 2 M_Z M_{Z'} \sz \cz \biggl(\cs + \frac{\sss}{\tanb} \biggr)\,,
\quad
\Gamma_{sZZ'} = 2 M_Z M_{Z'} \sz \cz \biggl(\sss - \frac{\cs}{\tanb} \biggr)\,,
\\
\Gamma_{hZ'Z'} &= 2 M^2_{Z'}\biggl(\cs \sz^2 - \frac{\sss}{\tanb}\cz^2 \biggr)\,,
\quad
\Gamma_{sZ'Z'} = 2 M^2_{Z'}\biggl(\sss \sz^2 + \frac{\cs}{\tanb}\cz^2 \biggr)
\esp
\label{eq:VSVV}
\end{equation}
where $\tanb = w/v$ is the ratio of the VeVs.
In the SWSM region of the parameter space the neutral boson mixing angle $\tZ$ 
is smaller than $\mathcal{O}\bigl(10^{-3}\bigr)$~\cite{Peli:2024fha}, 
hence we consider the leading contributions to the vertex factors $\Gamma_{SVV}$ 
in the limit $\tZ \to 0$ ($\cz=1,\: \sz=0$). 
\item Scalar-scalar couplings: 
\begin{equation}
\bsp
\Gamma_{hss} &= + \frac{M_H^2 + 2 M_s^2}{v}\, \sss \cs \left(\frac{\cs}{\tanb}-\sss\right) 
\,,\\
\Gamma_{shh} &= - \frac{2 M_H^2 + M_s^2}{v} \, \sss \cs \left(\frac{\sss}{\tanb}+\cs\right) 
\,.
\label{eq:VSSS}
\esp
\end{equation}
\end{itemize}
In these vertex factors the replacement $h\to s$ can be achieved by 
$(\cs \to \sss, \sss\to -\cs)$ and the replacement of one $Z\to Z'$ by {\em one factor} of 
$(\cz \to \sz, \sz\to -\cz)$. Thus, it is sufficient to define explicitly 
$\Gamma_{hZZ}$ and $\Gamma_{hss}$. 

\section{Production of scalar particles in the LHC}

The production cross sections of a scalar particle that mixes with the SM scalar, 
which we shall refer to as {\em Higgs-like new scalar}, can be computed based 
on Ref.~\cite{Djouadi:2005gi}. There are several production channels for the Higgs 
boson at the LHC, and similarly for the new scalar as well. These channels involve
(i) gluon-gluon fusion (ggF), 
(ii) associated production of the scalar with a vector boson (VS), 
(iii) vector boson fusion (VBF), and 
(iv) associated production of the scalar with a $t\bar{t}$ pair (ttS).

In the SWSM the new scalar is directly coupled to heavy quarks, so the dominant 
production channel is the gluon-gluon fusion, as in the SM.  The difference 
compared to the $\Gamma_{hf\bar{f}}$ vertex is only a proportionality factor 
obtained from the scalar mixing angle $\tS$ given in Eq.~\eqref{eq:VSff},
hence the gluon fusion cross sections for the production of the two scalar 
particles are proportional,
\begin{equation}
\sigma(gg \to h) = \cs^2  ~\sigma^{\text{SM}}(gg \to h)\,
\,,\quad
\sigma(gg \to s) = \sss^2 ~ \sigma^{\text{SM}}(gg \to h)|_{M_h \to M_s}
\,.
\end{equation}

The associated production of the scalar boson with a vector boson $V$ involves
the couplings \eqref{eq:VSVV} between the scalars and the vector bosons, which
provide factors of $\cs^2$ and $\sss^2$ as compared to the SM. In addition, 
there is a contribution due to the $SZ'$ channel, but it is negligibly small compared to the $SW$ or 
$SZ$ channels for small $Z'$ masses, $\xi = M_{Z'}/M_Z \ll 1$, relevant in the 
SWSM parameter space. The V-A couplings of the $Z$ boson to the quarks
also receives BSM corrections, but these are well measured quantities and the 
deviation from the SM must be small. Hence, the $Vh$ production cross section 
receives only an overall factor $\cs^2$ as compared to the SM model prediction, 
\begin{equation}
\sigma(pp \to Vh) = \cs^2 \sigma^{\text{SM}}(pp \to Vh)
\,.
\end{equation}
Thus, measuring $\sigma(pp \to Vh)$ and comparing it to the SM prediction for this process constrains $\cs$.

By far the most complicated process to compute its cross section is the vector 
boson fusion. However, the radiative corrections are known to be small 
\cite{Djouadi:2005gi}, so we consider this channel only at leading order (LO) 
in perturbation theory. The partonic process is $qq \to qq+ (V^* V^* \to h)$.
The squared matrix element of this process at LO is proportional to 
$G_F^3 M_V^8$, which means that the VBF process is also heavily suppressed, 
when $V=Z'$ and $\xi \ll 1$. Thus, only $V=Z$ and $W$ contributes with 
suppression factors $\cs^2$ for the Higgs production and $\sss^2$ for 
the new Higgs-like scalar production,
\begin{equation}
\sigma(qq \to qqh) = \cs^2 \sigma^{\text{SM}}(qq \to qqh)
\,,\quad
\sigma(qq \to qqs) = \sss^2 \sigma^{\text{SM}}(qq \to qqh)|_{M_h \to M_s}
\,.
\end{equation}

The scalar boson can also be produced in association with heavy quarks. 
This has the smallest contribution to the total production rate even when 
the heavy quark is the t-quark. The Higgs-like new scalar is directly 
coupled to heavy quarks just like in the Higgs boson, and the suppression 
factor in the production cross section is $\tan^2(\tS)$ just like in gluon-gluon fusion,
\begin{equation}
\sigma(pp \to t\bar{t}s) = \sss^2  \sigma^{\text{SM}}(pp \to t\bar{t}h)|_{M_h \to M_s}\,.
\end{equation}
Based on its small contribution to the total production cross section, we neglect this production channel in our analysis.

We see that all important production cross sections are proportional to the 
cross section of Higgs boson production in the SM 
with the relevant scalar mass value. Then it is sufficient to
precisely compute the Higgs boson production cross section for different 
values of the Higgs mass. One can use automated software for that purpose, 
or alternatively one can save data using plots from 
Ref.~\cite{ATLAStwiki}, which we did here to obtain the relevant $K$-factors.

\section{Decays of scalars}

\subsection{Total width of the Higgs boson }
\label{sect:h_width}
The SM theoretical prediction for the Higgs boson width is $\Gamma_h^{{\rm SM}} = 4.07~\MeV$, with a relative uncertainty of $4\%$~\cite{ParticleDataGroup:2024cfk}. The experimental measurements on the other hand are $\Gamma_h^{{\rm ATLAS}} = 4.5^{+3.3}_{-2.5}\,\MeV$~\cite{ATLAS:2023dnm}
and
$\Gamma_h^{{\rm CMS}} = 3.2^{+2.4}_{-1.7}\,\MeV$~\cite{CMS:2022ley},
display a much larger uncertainty than the SM theoretical prediction allowing for 
several BSM models to remain compatible with observations.

In the superweak and other U(1)$_z$ extensions with a light $Z'$ boson ($M_{Z'}\ll M_Z)$ the decay $h \to Z'Z'$ is allowed with partial width
\be 
\Gamma\bigl(h \to Z'Z'\bigr) = \frac{G_{\rm F} M_h^3}{16\sqrt{2}\pi}\left(\frac{\sss}{\tanb}\right)^2 +\mathcal{O}\left(\frac{M_{Z'}^2}{M_Z^2}\right),
\ee
and the total width of the Higgs boson is
\be 
\Gamma_h = \cs^2 \Gamma_h^{\rm SM} + \Gamma\bigl(h \to Z'Z'\bigr),
\ee
which can be used to constrain the parameters $\theta_S$ and $\tanb$. 
We see that the total decay width can be smaller in the extended model than that in the SM depending on the relative effect of the scalar mixing angle and the partial decay width of the Higgs boson into the new neutral gauge boson pair.

\subsection{Decay channels}
A Higgs-like new scalar has similar decay channels as the Higgs-boson as listed.

{\bf Fermionic:} The decays into SM fermions are only affected by an overall factor of  $\sss^2$ as compared to the SM. If the mass of the Higgs-like scalar is above the $2m_N$ threshold, it can also decay into a pair of HNLs, with partial with
\beq 
\Gamma\bigl(S \to N_i N_i \bigr) = \frac{G_{\rm f} M_s m_{N_i}^2}{8\sqrt{2}\pi}\biggl(1-\frac{4 m_{N_i}^2}{M_s^2}\biggr)^{3/2}\biggl(\frac{\cs}{\tanb}\biggr)^2
\,.
\label{eq:StoNN}
\eeq

The Higgs boson can also decay into an HNL pair, provided the process is 
allowed kinematically. The corresponding partial decay width can be obtained 
form formula \eqref{eq:StoNN} with the replacements $M_s \to M_h$ and $\cs \to \sss$.

{\bf Loop induced:} This category includes the decays into $\gamma\gamma$, 
$gg$ and $Z\gamma$ as well as $Z'\gamma$. In the important decay channels 
$s\to\gamma\gamma$ and $gg$, the decay widths are only multiplied by an overall $\sss^2$ factor,
\begin{equation}
\Gamma(s \to \gamma\gamma,\:Z\gamma,\:gg) = \sss^2 \Gamma^{\text{SM}}(h \to \gamma\gamma,\:Z\gamma,\:gg)|_{M_h\to M_s}
    \,.
\end{equation}

The $Z'\gamma$ channel is also interesting because the LHC cast 
exclusion limits specifically on this channel (with branching ratio ${\rm Br}(Z'\gamma) < 2\,\%$), 
but the prediction for this branching fraction is much smaller then the 
experimental limit. In the case of a light $Z'$ and a modest $Z-Z'$ mixing, 
its contributions are negligibly small.

{\bf Decays into a pair of heavy vector bosons:} The decays of the 
Higgs-like scalar into a charged $W$-pair is only affected by an overall factor 
of $\sss^2$ as compared to the SM prediction. Decays into heavy neutral bosons 
are controlled by the $S Z^{(\prime)}Z^{(\prime)}$ vertex. The vertex $S Z Z'$ 
is naturally suppressed, but the $S Z' Z'$ may be large.

{\bf Decays into scalars:} In case $M_s > 2 M_h$, the channel $s \to hh$ 
opens up, while the opposite decay, $h\to ss$ is  excluded by the results for
$\Gamma_h^{\rm exp.}$ compared to the SM prediction.

The decay properties of a Higgs-like new scalar depend on all free parameters
of the model. An indirect restriction on the parameter space can be derived from
the experimental constraints on the Higgs-boson width $\Gamma_h$.

A more refined analysis reveals that for $\xi \ll 1$, such as the case in the SWSM, 
the free parameters reduce to a set of four parameters \cite{Peli:2023fyb}:
\beq 
M_s,~\sss,~\tanb,~\text{and}~m_{N_i},
\eeq
where 
$\tanb$ absorbs the free parameters of the gauge sector 
in the combination
\beq 
\label{eq:tanb_mzp}
\tanb \propto \frac{M_{Z'}}{|\sz|M_Z}\,. 
\eeq
The exact formula contains an additional factor depending on the 
$z$-charge of the BEH field and the kinetic mixing between the U(1) 
gauge fields \cite{Peli:2024fha}, which is not important in our 
analysis.  We note that for the B-L $U(1)$ extension the 
proportionality factor is zero, so the scalar and gauge sector 
parameters are not related in that model.

\section{Implementation}

We computed the Higgs boson production cross section 
$\sigma\bigl(pp \to h +X\bigr)$ at $\sqrt{s} = 13.6~\TeV$ 
center of mass energies in proton proton collisions for several values of $M_h$ in the range [$100~\GeV, 1~\TeV$]. 
Producing a Higgs-like scalar means that this cross section is multiplied with $\sss^2$.
In principle, the $Z'$ boson affects these production rates in a nontrivial 
way, but these effects are negligibly small for $M_{Z'} \ll M_{Z}$, which is 
relevant for the SWSM.

We used the NNPDF3.0 set for the parton density functions (PDF), the LO 
set for LO predictions and the NLO set for the computations at the 
next-to-leading order (NLO) accuracy. We use the running strong coupling 
from the chosen PDF set. We take the values of the other input parameters
$m_b^{\rm pole}$, $m_t^{\rm pole}$, $G_f$, $M_Z$ and $M_W$ from 
Ref.~\cite{ParticleDataGroup:2024cfk}. 

We compute the production cross section from {ggF} process at NLO QCD  with NNLO QCD and NLO EW corrections included as a K-factor. The {Vh} process is implemented at NLO QCD and finally we compute the {VBF} at LO as the NLO corrections are known to be small with $K^{\rm VBF}_{\rm NLO} \sim 1.1$.

We include a {\em Mathematica} notebook \texttt{swsm\_scalar.nb} 
which contains a precomputed set of the scalar boson productions cross sections and precisely computes the decay rates of the Higgs boson and the Higgs-like scalar.

\section{Signal strengths}

If a light neutral vector boson $Z'$ exists, the Higgs boson can decay 
into a $Z'Z'$ pair, which affects the signal strengths in other channels. 
The signal strength measured at the LHC is 
\be 
\mu_{\rm exp.} = \frac{\bigl(\sigma\, {\rm Br}\bigr)_{\rm obs}}{\bigl(\sigma\, {\rm Br}\bigr)_{\rm SM}}\,,
\ee
while the theoretical value in the presence of new physics is 
\be 
\mu_{\rm th.} = \frac{\bigl(\sigma\, {\rm Br}\bigr)_{\rm BSM}}{\bigl(\sigma\, {\rm Br}\bigr)_{\rm SM}}
\ee
where the BSM subscript refers to the prediction in the full BSM model, 
including the SM contributions. The most precisely measured channel is 
$h\to Z Z^*$~\cite{ParticleDataGroup:2024cfk}. In the superweak model and in 
general when the Higgs boson cannot decay into an $ss$ or HNL pair, the 
predicted signal strength in the $ZZ$ channel is 
\be 
\mu_{ZZ} = \frac{\cs^4 \Gamma_h^{{\rm SM}}}{\cs^2 \Gamma_h^{{\rm SM}} + \Gamma\bigl(h \to Z' Z'\bigr)}
\ee
because $\sigma_h^{{\rm SWSM}} = \cs^2 \sigma_h^{{\rm SM}}$ and
$\Gamma^{\rm SWSM}(h \to ZZ) = \cs^2\Gamma^{\rm SM}(h \to ZZ)$. 
Using the PDG24 \cite{ParticleDataGroup:2024cfk} values 
$\bigl(\Gamma_h \simeq 0.00407 ~\GeV, M_h \simeq 125~\GeV\bigr)$, we obtain 
\be 
\label{eq:hzz_signal}
\mu_{ZZ} = \frac{\cs^4 }{\cs^2 + 78.74 \bigl(\sss / \tanb \bigr)^2}\,,
\ee
which approaches $\cs^2$ for large values of $\tanb$. In that case the 
total signal strength provides a considerably more severe limit. 
The latter sums over all production and decay channels, hence 
\be 
\mu_{\rm tot} = \frac{\sigma_h^{{\rm BSM}}}{\sigma_h^{{\rm SM}}} = \cs^2,
\ee
in which case the experimental values are 
\cite{ATLAS:2022vkf,CMS:2022dwd} 
\be 
\mu^{\rm ATLAS}_{\rm tot} = 1.05 \pm 0.06, \quad \quad
\mu^{\rm CMS}_{\rm tot} = 1.02 \pm 0.06.
\ee
Then, the total signal strength yields the bounds independent of $\tan\beta$
\be 
\theta^{\rm ATL}_{S,{\rm tot}} = 0.27,
\quad\quad
\theta^{\rm CMS}_{S,{\rm tot}} = 0.32.
\ee

\section{Benchmarks and exclusion bounds}

\begin{figure}[t]
\centering
\includegraphics[width=0.7\linewidth]{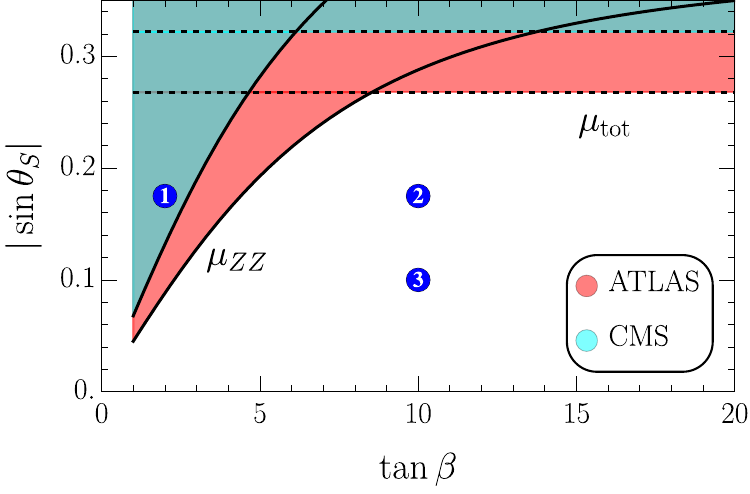}
\caption{\label{fig:signal_bounds} 
Limit on the sine of the scalar mixing angle $\theta_S$ as a function of $\tan\beta$. The 
red (cyan) region is excluded at $95\%$ C.L. 
by the ATLAS (CMS) measurement
of the total and $h\to ZZ^*$ signal strengths. The blue disks with numbers $i$ correspond to benchmark points BP$i$.
}
\end{figure}

For sufficiently small values of $\tanb$ the bound obtained from $\mu_{ZZ}$ 
tends to be more constraining than that obtained from $\mu_{\rm tot}$. 
The experimentally measured values for the $ZZ^*$ signal strength are \cite{ATLAS:2022vkf, CMS:2022dwd}
\be 
\mu^{\rm ATLAS}_{ZZ} = 1.04 \pm 0.09 \,, \quad \quad
\mu^{\rm CMS}_{ZZ} = 0.97 \pm 0.12 \,,
\ee
which includes all the productions channels. Figure~\ref{fig:signal_bounds} shows 
limits obtained from the total signal strength measurements as well as those from the $ZZ^*$ channel for $\tan\beta\leq20$. The 
colored region 
on the $\sss-\tanb$ plane is 
excluded at $95\%$ confidence level (C.L.).
The blue discs with numbers $i$ correspond to benchmark points BP$i$ we propose 
based on the exclusion bounds obtained from the signal strength measurements.
Explicitly, the three benchmark points are:
\be
\bsp\label{eq:benchmark_def}
{\rm BP1}: \tS &= 0.175, \tanb = 2, M_S = 500\,\GeV \,,
\\
{\rm BP2}: \tS &= 0.175, \tanb = 10, M_S = 500\,\GeV \,,
\\
{\rm BP3}: \tS &= 0.10, \tanb = 10, M_S = 1000\,\GeV \,.
\esp
\ee
where BP1 is allowed by $\mu_{\rm tot}$, but excluded by $\mu_{\rm ZZ}$, while BP2 and BP3 are
still allowed. 
BP1 is chosen such that  
finding a scalar corresponding to BP1 would exclude 
a large class of BSM models predicting a light $Z'$ boson.
Using Eq.~\eqref{eq:tanb_mzp} the gauge
sector parameters are
\be
\bsp
{\rm BP1}: M_{Z'} &= 18\,\MeV, \quad\sz = 10^{-4}  \,,
\\
{\rm BP2}:  M_{Z'} &= 91\,\MeV, \quad \sz = 10^{-4}  \,,
\\
{\rm BP3}:  M_{Z'} &= 91\,\MeV, \quad \sz = 10^{-4}  \,.
\esp
\ee

The Higgs-portal coupling $\lambda$ can be expressed via the relation
\be 
\lambda = \frac{M_S^2 - M_H^2}{v^2}\,\frac{\sss\cs}{\tanb } \,,
\ee
yielding 
\be 
\lambda_{\rm BP1} = 0.33, \quad
\lambda_{\rm BP2} = 0.07
\quad\text{and}\quad
\lambda_{\rm BP3} = 0.16\,.
\ee
The SWSM model contains three families HNLs $N_i$. We consider the 
masses of these particles $M_{N_1} = \mathcal{O}(M_{Z'})$ and
$M_{N_2}=M_{N_3} = 100\,\GeV$ for all benchmark points. Then, the HNLs do
not contribute to the decays of the Higgs boson, and can explain dark
matter abundance \cite{Iwamoto:2021fup}. We present the productions
cross sections and branching ratios corresponding to BP1, BP2 and BP3
in Table~\ref{tab:benchmark_data}.
Further points can be obtained similarly using the {\em Mathematica} code available on request.
\begin{table}[t] 
\centering
\caption{The productions cross sections and branching fractions corresponding to the benchmark points defined in Eq.~\eqref{eq:benchmark_def}.\label{tab:benchmark_data}}
\renewcommand{\arraystretch}{1.05}
\begin{tabularx}{0.75\textwidth}{c||r|r|r||r|r|r} 
\toprule
 & \multicolumn{3}{|c||}{$h$} & \multicolumn{3}{|c}{$s$} \\
\midrule
 -- & BP1 & BP2 & BP3 & BP1 & BP2 & BP3 \\ [0.5ex] 
\hline
$\sigma_{\rm prod}$ [pb] & $53$ & $53$ & $54$ & $0.16$ & $0.16$ & $0.001$ \\ 
\hline
$\Gamma$ [$\GeV$] & $6.4\cdot 10^{-3}$ & $4.0\cdot 10^{-3}$ & $4.0\cdot 10^{-3}$ & $8.3$ & $2.9$ & $8.5$ \\
\hline
${\rm Br}\!\left(hh\right)$  & -- & -- & -- & $0.09$ & $0.23$ & $0.20$  \\
\hline
${\rm Br}\!\left(W^+W^-\right)$  & $0.13$ & $0.21$ & $0.21$  & $0.13$ & $0.37$ & $0.37$ \\
\hline
${\rm Br}\!\left(ZZ\right)$  & $0.016$ & $0.026$ & $0.026$ & $0.06$ & $0.18$ & $0.18$ \\
\hline
${\rm Br}\!\left(Z'Z'\right)$  & $0.38$ & $0.024$ & $0.008$ & $0.6$ & $0.07$ & $0.19$ \\
\hline
${\rm Br}\!\left(b\overline{b}\right)$  & $0.36$ & $0.57$ & $0.58$ & $<10^{-4}$ & $<10^{-4}$ & $<10^{-4}$ \\
\hline
${\rm Br}\!\left(t\overline{t}\right)$  & -- & -- & -- & $0.05$ & $0.14$ & $0.05$ \\
\hline
$2\times{\rm Br}(N N)$  & -- & -- & -- & $0.07$ & $0.009$ & $0.07$ \\
\bottomrule
\end{tabularx}
\end{table}
\section{Summary}

In our analysis we focused on U(1) extensions of the SM where the new gauge
boson becomes massive via the spontaneous breaking of the new $U(1)$ symmetry
by a complex scalar field, such as in the superweak extension of the SM.
We have shown that in the case of a light $Z'$, the new decay channel 
$h \to Z'Z'$ alters the signal strengths in all decay channels 
of the Higgs boson.
Here we focused on the $ZZ^*$ channel, which is measured experimentally with high 
precision. We obtained bounds on the free parameters of the model: the Higgs -- 
new scalar mixing angle $\tS$ and the ratio $\tanb$ of the VeV of the new scalar 
to that of the BEH field. We presented three benchmark points -- one (BP1) already 
excluded by the signal strength measurement in the $ZZ^*$ decay channel of the
Higgs boson and two other ones (BP2, BP3) still allowed. Finding a new scalar $s$
corresponding to BP1 would mean that no new light $Z'$ boson exits 
as it would violate the exclusion bound derived from the effect of the decay $h \to Z'Z'$.
Further
benchmark points can be generated easily using the {\em Mathematica} 
notebook available on request.

\acknowledgments{This research was funded by Excellence Programme of the Hungarian Ministry of Culture and Innovation grant number TKP2021-NKTA-64, by the National Research, Development and Innovation Office grant number NKFI-150794, and by the Hungarian Scientific Research Fund grant number PD-146527.}


\end{document}